\documentclass[prb,floatfix,twocolumn,superscriptaddress]{revtex4}
\usepackage{graphics,epsfig,amsfonts,amssymb,amsmath}%

\begin{document}
\title{THEORY OF ONE AND TWO DONORS IN SILICON}
\author{A. L. Saraiva}
\email{also@if.ufrj.br}
\affiliation{University of Wisconsin-Madison, Madison, Wisconsin 53706, USA}
\affiliation{Instituto de Fisica, Universidade Federal do Rio de Janeiro, Caixa Postal 68528, 21941-972 Rio de Janeiro, Brazil}
\author{A. Baena}
\affiliation{Instituto de Fisica, Universidade Federal do Rio de Janeiro, Caixa Postal 68528, 21941-972 Rio de Janeiro, Brazil}
\author{M. J. Calder\'{o}n} 
\affiliation{Instituto de Ciencia de Materiales de Madrid, ICMM-CSIC, Cantoblanco, E-28049 Madrid, Spain}
\author{Belita Koiller}
\affiliation{Instituto de Fisica, Universidade Federal do Rio de Janeiro, Caixa Postal 68528, 21941-972 Rio de Janeiro, Brazil}
\date{\today}

\begin{abstract}
We provide here a roadmap for modeling silicon nano-devices with one or two group V donors (D). We discuss systems containing one or two electrons, that is, D$^0$, D$^-$, D$_2^+$ and D$_2^0$ centers. The impact of different levels of approximation is discussed. The most accurate instances -- for which we provide quantitative results -- are within multivalley effective mass including the central cell correction and a configuration interaction account of the electron-electron correlations. We also derive insightful, yet less accurate, analytical approximations and discuss their validity and limitations -- in particular, for a donor pair, we discuss the single orbital LCAO method, the H\"uckel approximation and the Hubbard model. Finally we discuss the connection between these results and recent experiments on few dopant devices.
\end{abstract}

%
%
\maketitle

\section{Introduction}
\label{intro}
In a letter to Peierls in 1931, Pauli stated that semiconductors ``{\it are a filthy mess}''~[\onlinecite{lettertopeierls}]. Curiously, the impurities and defects to which Pauli refers led the way to the revolution severely impacting industry and society over the last many decades, and still under way. Development and advances in  semiconductor-based devices started less than 20 years after Pauli's statement, as the transistor operated for the first time in 1947. This achievement  prompted intensive research  for a deeper understanding of semiconductors and the role of dopants. Currently, as ``Moore's Law'' approaches its limit with devices  reaching the atomic scale, quantum behavior of electrons and spins drive the advancement of the semiconductor research into the new fields of quantum electronics and spintronics~[\onlinecite{sciencereview,rmpreview,morley2014}].

Doping allows controlling the sign and density of carriers in a semiconductor, a flexibility not achievable in conductors or in insulators. For the  applications where current carriers perform the needed operations, donor electrons are promoted to the nearby conduction band and similarly holes from acceptors operate in the valence band. In this case the concentration of donors and acceptors and their binding energies are the only relevant quantities defining the behavior of the doped material. A macroscopic concentration of dopants leads to macroscopically observable currents or voltages, and at this point the quantum behavior becomes irrelevant.

We revisit here the theoretical treatment of donors D in semiconductors from a single substitutional donor perspective, considering that its active electron(s)  remains bound to the core, and similarly for donor pairs D$_2$, a scenario where full quantum behavior prevails. Special attention must be paid to the electronic bound state wavefunctions.

This paper is organized as follows. In Sec.~\ref{hydro} we briefly review the effective mass theory (EMT) for shallow donors, highlighting its successes and limitations. In Sec.~\ref{multi}, multivalley and anisotropy effects are analysed, details of the approach used are given, and the results for neutral donors are shown in Sec.~\ref{dzero}. Section~\ref{dminus} reports on a donor with two bound electrons, constituting a D$^{-}$ negatively charged donor. The donor pair is studied in Secs.~\ref{d2plus} and~\ref{d2} for the one and two electron states respectively. Sec.~\ref{sec:discussion} discusses simplified models for the two donor problem and makes connections with experiments. Finally, summary and conclusions are presented in Sec.~\ref{sec:summary}. In order to keep the text reasonably self-contained, well established basic material and results are included with proper references  provided. 

 \section{Hydrogenic model for donors in semiconductors}
 \label{hydro}

The single donor description is one of the simplest and most successful implementations of  the EMT. It is based on the usual effective-medium assumptions, namely that:
     \begin{enumerate}
  \item  {\it the perturbation potential due to a substitutional donor varies slowly at the scale of the lattice constant of the host semiconductor;}

  \item   {\it  the envelope functions of the bound states extend over long enough distances such that it is composed by a very narrow distribution of Bloch wavevectors {\bf k} around the band minimum.}

   \end{enumerate}
Further assuming that the conduction band lower edge is non-degenerate and isotropic, one obtains the hydrogenic atom hamiltonian
\begin{equation}
{\mathcal H}_H=-\frac{\hbar^2\nabla^2}{2m^*}-\frac{e^2}{4\pi\epsilon r}
\label{eq:hydro}
\end{equation}
with the electron mass substituted by the effective mass $m^*$ and the Coulomb potential screened by the static dielectric constant $\epsilon$. This gives a hydrogenic atom of effective Bohr radius $a^* = a_0\epsilon/m^*$ and ground state binding energy $E^* = E_H m^*/\epsilon^2$, with $a_0=0.053$ nm and  $E_H=13.6$~eV the respective values for Hydrogen in vacuum.
The wavefunction for the ground state electron is then the product of the hydrogenic envelope-function by the band-edge Bloch function
\begin{equation}
\Psi = \frac {1}{\sqrt{\pi {a^*}^3}}e^{-r/a^*} e^{i{\bf k \cdot r}}u_{\bf k}({\bf r}),\label{eq-psih}
\end{equation}

\begin{figure}
\leavevmode
\includegraphics[clip,width=0.48\textwidth]{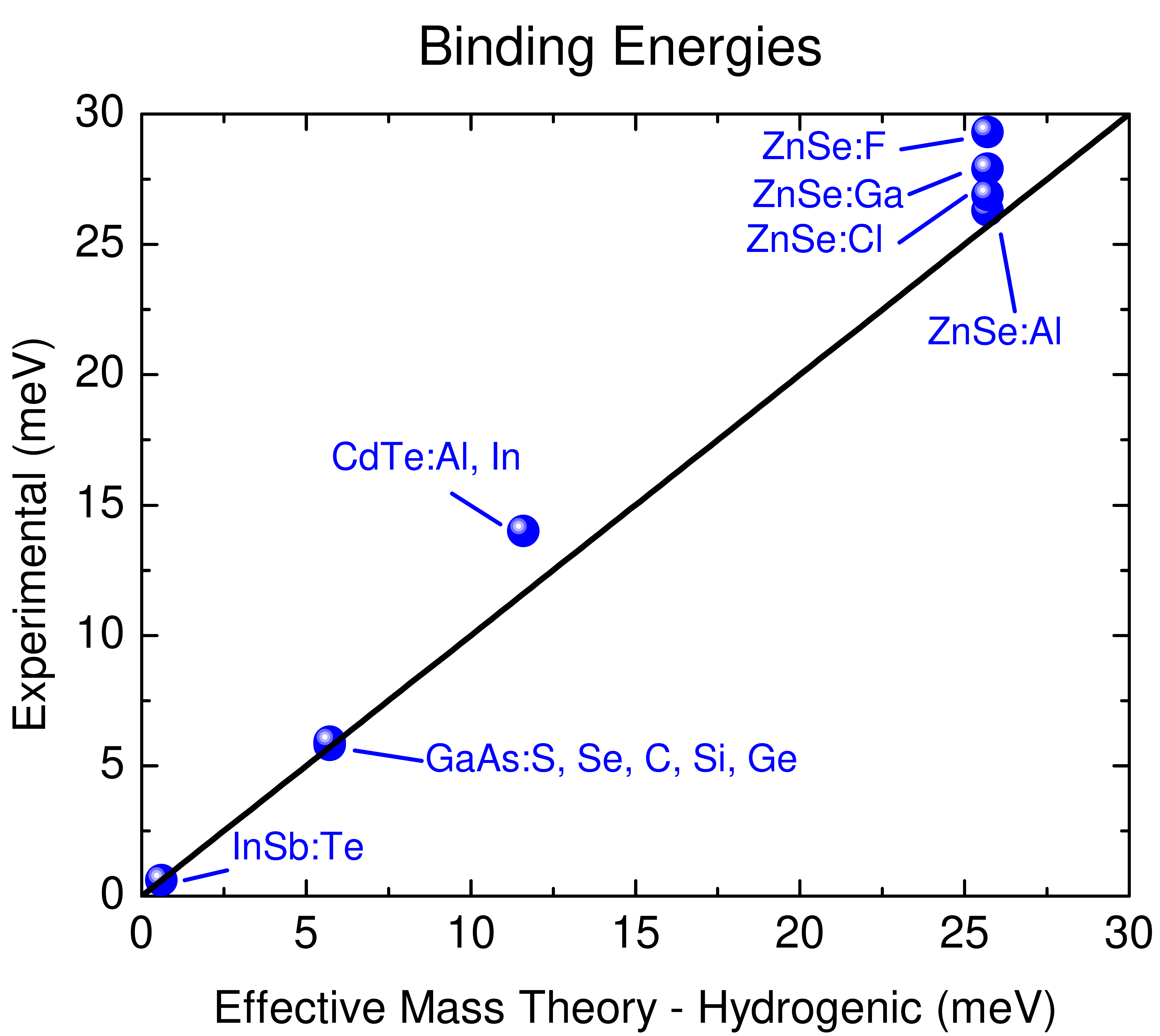}
\caption{Comparison of the experimental donor binding energies for single valley materials (blue dots) with those calculated from the effective mass hydrogenic expression (black line). The agreement is very good mainly for the less bound donors. The ground state energies for donors in silicon (not shown here) are off the given scale and present a large variation ($\sim50$\%) as a function of the donor species, see Table~\ref{radii}. The same spread of values is found for germanium~[\onlinecite{ramdas1981}].}
\label{fig1}
\end{figure}

The EMT provides a highly accurate description of shallow donors in single valley materials. It involves no fitting parameters, as the  binding energy is dependent on the material only through $\epsilon$ and $m^*$, leading to a value of $E^*$ that is independent of the donor species. This is in fact clearly demonstrated experimentally for example in GaAs, where deviations from the hydrogenic expression are hard to detect~[\onlinecite{Fetter71}], as illustrated in Fig.~\ref{fig1}. In contrast, the binding energy of donors in Si and Ge is strongly dependent on the donor species~[\onlinecite{ramdas1981}].

\section{Multivalley effects}
\label{multi}

The case of multivalley semiconductors is not as simple. In a recent work~[\onlinecite{salfi2014}], spatially resolved spectroscopy of isolated As donors in silicon was combined with multiband semi-empirical tight binding theory to study the valley interference. These results provide direct evidence for the rich valley structure predicted over 60 years ago by Kohn and Luttinger~[\onlinecite{KL}] (KL) and further outline the role of the environment on valley repopulation.

With the $N$ minima (valleys) at finite wavevectors, the symmetry of the wavefunction is significantly lower, and highly anisotropic effective masses are not uncommon -- in silicon the longitudinal and transverse effective masses are $m_L$= 0.916 $m_e$ and $m_T$=0.191 $m_e$ at the $N=6$ minima. The hydrogenic hamiltonian Eq.~(\ref{eq:hydro}) may be adapted to account for the mass anisotropy, as discussed in Sec.~\ref{uncoupled}.

The mere reconciliation with the mass anisotropy still does not lead to results compatible with the experimental binding energies. The missing ingredient is the elusive valley-orbit coupling. In the absence of valley-orbit coupling, a wavefunction equivalent to Eq.~(\ref{eq-psih}) is obtained for each of the $N$ valleys, leading to an $N$-fold degenerate ground state.

The $1/r$ Coulomb potential singularity at $r=0$ is clearly incompatible with the assumptions discussed in Sec.~\ref{hydro} -- it actually mediates a finite coupling among the $N$ valleys. This accounts for over 30\% of the binding energy in silicon donors, as will be discussed in Sec.~\ref{vo}.  We briefly present some of the most successful semi-empirical approaches to describe these departures from the hydrogenic model, namely the first order perturbation intervalley coupling (Sec.~\ref{intervalley}) and the central cell correction (Sec.~\ref{cc}). A theoretical description of valley-orbit coupling from first principles is still lacking.

\subsection{Uncoupled anisotropic valleys \label{uncoupled}}
The first successful approach to treat donors in Si was presented by KL~[\onlinecite{KL}] (which also set stronger formal basis for the EMT) and Kittel and Michel~[\onlinecite{KM}]. Taking initially the EMT assumption (ii) to be valid -- namely that only small deviations from the wavevector at the band minimum ${\mathbf k}_{min}$ contribute to the wavefunction-- means that any individual valley does not couple to the other five. In this way each valley is solved independently and the problem regains the hydrogenic simplicity.

The band anisotropy leads to a specific  effective mass hamiltonian for the different valleys. For instance, for the 2 valleys along $z$, it reads
\begin{equation}
{\mathcal H}_z= -\frac{\hbar^2}{2 m_{T}}\left(\frac{\partial^2}{\partial x^2} + \frac{\partial^2}{\partial y^2}\right) - \frac{\hbar^2}{2 m_{L}}\frac{\partial^2}{\partial z^2}+V(r).
\label{equation3}
\end{equation}
The ground state is found variationally, choosing a trial function for the ${\mathbf k}_z = 0.85 \frac {2\pi}{a_{\rm Si}} (001) $ valley ($a_{\rm Si}$ is the Si conventional lattice parameter) of the shape of a deformed 1s orbital with two radii $a$ and $b$~[\onlinecite{foot}]
\begin{equation}
\phi_z({\mathbf r}) = \frac{1}{\sqrt{\pi {a^2 b}}}
e^{-{\sqrt{(x^2+y^2)/a^2 + z^2/b^2}}} e^{i{\bf k \cdot r}}
u_{\bf k}({\bf r}).\label{eq-psikl}
    \end{equation}
For other valleys at $\{{\mathbf k_\mu}\}, \mu=\{x,-x,y,-y,z,-z\}$, the hamiltonians ${\mathcal H}_\mu$ and the wavefunctions $\phi_\mu$ are immediately obtained exchanging $z$ by $x$ or $y$ accordingly. The deformed orbital Bohr radii $a$ and $b$ are taken as variational parameters to minimize the expectation value of $\langle\phi_\mu|{\mathcal H}_{\mu}|\phi_\mu\rangle$. For the hydrogenic potential $V(r)=V_{H}(r)=-e^2/4\pi\epsilon_{\rm Si} r$, the ground state is sixfold degenerate, and one recovers the KL result with energy $E_{\rm KL}=-31.2$ meV, and an anisotropy $b/a = 0.56$.

\subsection{Valley-orbit coupling \label{vo}}

The binding energy is significantly underestimated if the valley-orbit coupling is disregarded, as can be seen by comparison of $E_{\rm KL}$ to the observed values for shallow donors in Si in Table~\ref{radii}. The valley-orbit coupling breaks the sixfold degeneracy, leading to a significant reduction of the ground state energy.

\begin{figure}
\leavevmode
\includegraphics[clip,width=0.48\textwidth]{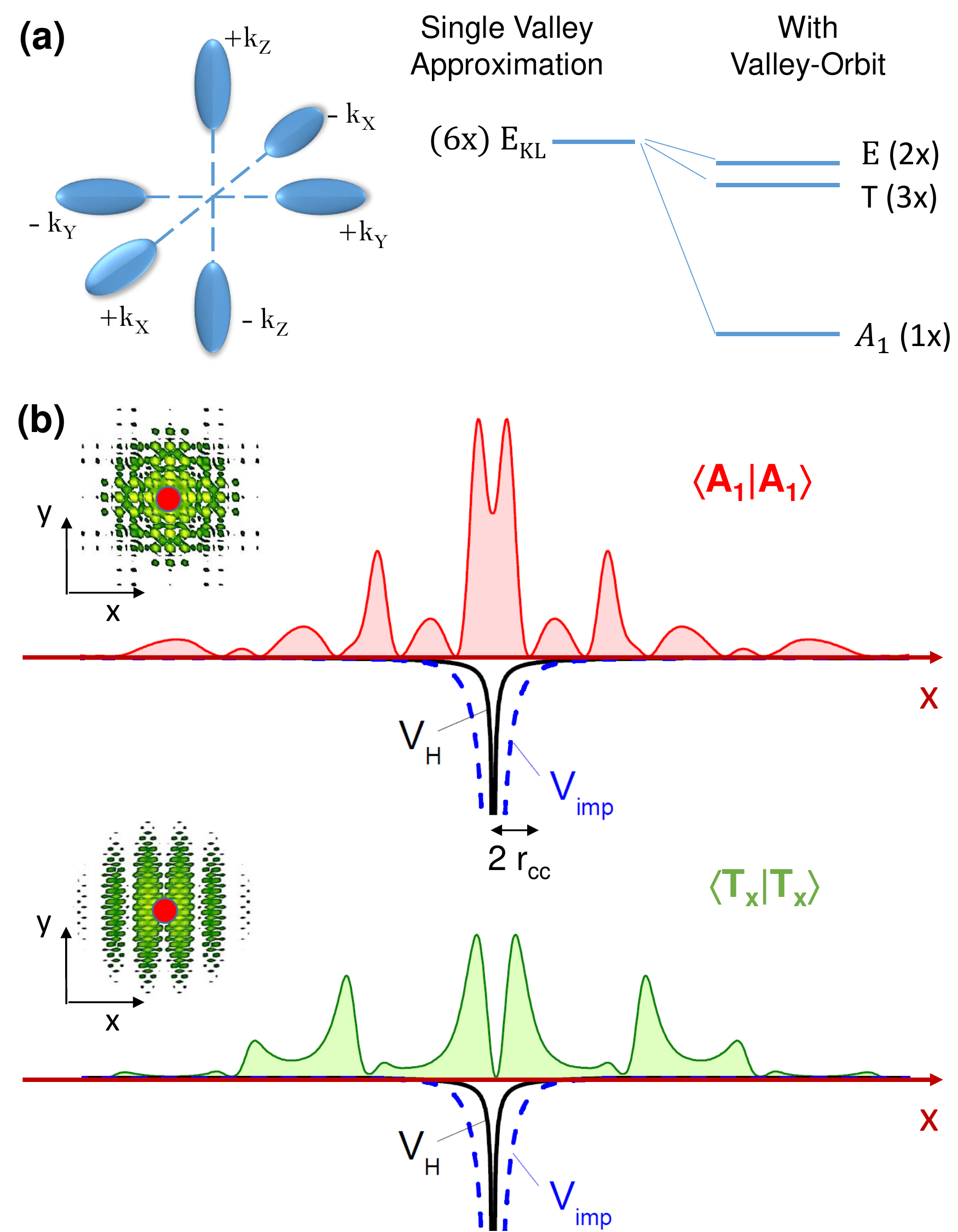}
\caption{(a) Representation in k-space of the six constant energy surfaces around the degenerate conduction band minima in Si. (b) Splitting of the sixfold degenerate conduction band minimum by a tetrahedral crystal environment combined with the singular donor potential. (c) Charge distribution for the A$_1$ and $T_x$ states along the $x$-axis (large) and at the (001) Si crystal plane (inset). Note that A$_1$ has significant charge density at the donor site, while the T$_x$ state has a node. The lower lines curving down give the plain screened Coulomb potential $V_H$ (solid line) and the central-cell corrected potential $V_{\rm imp}$ (dashed line) which is more attractive for distances from the donor nucleus  smaller than $\sim r_{cc}$. Due to the larger charge density of the A$_1$ state within $r_{cc}$, its energy is well below the T$_2$ and E states.}
\label{fig2}
\end{figure}

Even if the strength of this coupling is not known, group theory arguments
give the appropriate superposition of valley states induced when the
spherically symmetric potential is disrupted by the tetrahedral crystal field
potential. In other words, the irreducible representation associated with the
ground state of a spherical potential with six-fold valley degeneracy (the 1s
manifold) becomes reducible in the presence of a local tetrahedral crystal
field: the sixfold degenerate level splits into states that have the symmetry
of the different irreducible representations of the $T_d$ group. In this
case, this leads to a singlet with A$_1$ symmetry, a triplet with T$_2$
symmetry and a doublet with $E$ symmetry. Each of the six states in the 1s
manifold correspond to a particular combination of the envelope-modulated Bloch functions from the
six degenerate valleys ${\phi_\mu}$ illustrated in Fig.~\ref{fig2}. These states are given
explicitly by~[\onlinecite{Kohn_review}]

\begin{widetext}
\begin{eqnarray}
\Psi_{A_1}({\bf r})&= & \frac{1}{\sqrt{6}}\left[\phi_x ({\bf r}) +\phi_{-x} ({\bf r}) +\phi_y ({\bf r}) +\phi_{-y} ({\bf r})+\phi_z ({\bf r}) +\phi_{-z} ({\bf r})\right] \nonumber \\
\Psi_{T_2^x}({\bf r})&= &  \frac{1}{\sqrt{2}}\left[\phi_x ({\bf r}) -\phi_{-x} ({\bf r}) \right] \nonumber \\
\Psi_{T_2^y}({\bf r})&= &  \frac{1}{\sqrt{2}}\left[\phi_y ({\bf r}) -\phi_{-y} ({\bf r}) \right] \nonumber \\
\Psi_{T_2^z}({\bf r})&= & \frac{1}{\sqrt{2}}\left[\phi_z ({\bf r}) -\phi_{-z} ({\bf r}) \right] \nonumber \\
\Psi_{E^z}({\bf r})&= & \frac{1}{\sqrt{12}} \left[\phi_x ({\bf r}) +\phi_{-x} ({\bf r}) +\phi_y ({\bf r}) +\phi_{-y} ({\bf r})-2\phi_z ({\bf r}) -2\phi_{-z} ({\bf r})\right] \nonumber\\
\Psi_{E^{xy}}({\bf r})&= & \frac{1}{2}\left[\phi_x ({\bf r}) +\phi_{-x} ({\bf r}) -\phi_y ({\bf r}) -\phi_{-y} ({\bf r})\right].
\label{eq:single-donor-wf}
\end{eqnarray}
\end{widetext}

\begin{figure}
\leavevmode
\includegraphics[clip,width=0.48\textwidth]{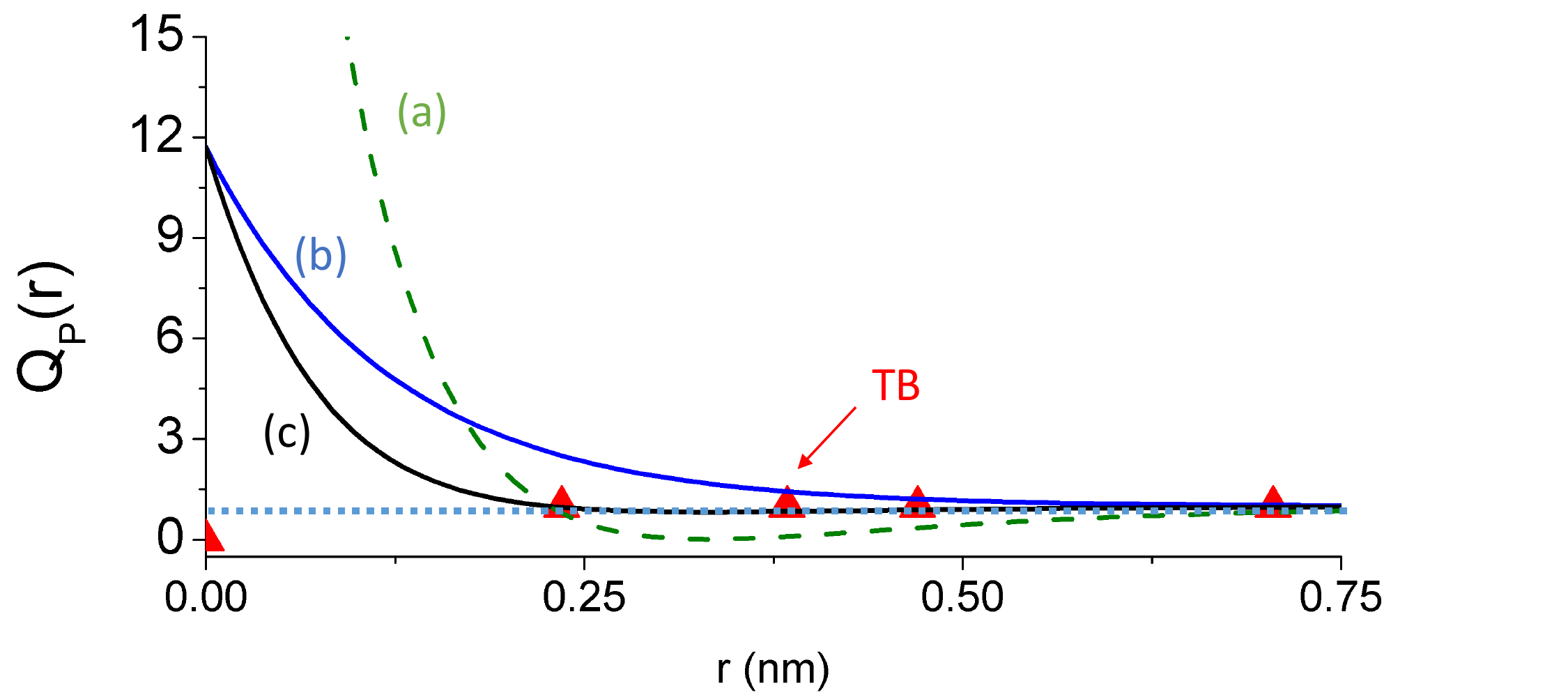}
\caption{
Comparison between different models adopted in the literature for the central cell correction. The figure gives $Q_P(r)$ such that $V_{\rm P}(r) = -Q_P(r) V_H(r)$ calculated for  P in Si. The different curves correspond to (a) Ref.~[\onlinecite{wellard2005}], (b) $V_{\rm imp}(r)$ as in Eq.~(\ref{eqrcc}), (c) Ref.~[\onlinecite{nara1966}], and the triangles give the approximation used in tight-binding~[\onlinecite{Martins05}]. The limit of a point charge $Q_P=1$ is shown as a dotted horizontal line.  Note that, for the tight-binding, $Q_P(r)=1$  for all sites except the impurity site where $V_{\rm imp}$  is taken as a constant. In this case we indicate $Q_P(r)=0$ in the figure, which  eliminates the $1/r$ divergence at $r\to 0$, but a constant $U_0$ should be added.
}\label{fig3}
\end{figure}

The direct calculation of the valley-orbit coupling is not trivial, though.
We are unaware of any successful description of the complete spectrum of the
1s manifold from first principles. This difficulty is due to the incomplete
knowledge regarding the impurity potential. The picture of a $1/r$ point
charge potential screened by a dielectric constant $\epsilon_{\rm Si}$ is
realistic only at large distances from the impurity site. Close to the
nucleus, the dielectric response becomes inhomogeneous. Furthermore, the
valence core shells for the impurity are different from the Si atoms,
presenting a strong departure from the excess proton picture. Atomic species
differ from each other at the core region, leading to the strong donor
species dependence of the binding energy in silicon in contrast with other
semiconductors. Early attempts of correcting the Coulomb potential
adopting the \textit{ab initio} dielectric function were modestly successful for the isovalent
impurity P and were inaccurate for all other group V donors (for a complete discussion of early works,
see Ref.~[\onlinecite{pantelides1974}]). Successful
attempts involve empirical fitting ingredients~[\onlinecite{nara1966,Martins05,wellard2005}]. Some examples used in
the literature to describe this departure from the point charge picture are
shown in Fig.~\ref{fig3}. Two among the main strategies for describing
the valley-orbit induced splitting are described next.

\subsection{First order perturbation theory: the intervalley coupling \label{intervalley}}

The EMT assumptions (i) and (ii) are clearly incompatible with the Coulomb
potential, which is singular at the donor site $r=0$, implying that different
valley states may couple. The single-valley KL description neglects this singularity, i.e. ${\mathcal H}$ defined in Eq.~(\ref{equation3}) does not include the singular part of the Coulomb potential $V_H$.

We may then include the intervalley coupling as a perturbation to the KL description. The total Hamiltonian would be $H={\mathcal H}+H'$ with $H'$ accounting for the singular part of the potential around the core region $r \to 0$, i.e. $V_H(r\to 0) $ plus additional corrections as discusssed below. For simplicity in notation we are assuming an isotropic approximation (further discussed in Sec.~\ref{dzero}) which implies that ${\mathcal H}_z= {\mathcal H}_x = {\mathcal H}_y = {\mathcal H}$. Now, the single-valley approximation implies $\langle
k_\mu|{\mathcal H}|k_\nu \rangle = E_{\rm KL}\delta_{\mu,\nu}$, where
$|k_\mu\rangle$ stands for $\phi_\mu$. The valley-coupling part, $H'$, is
added and treated following standard degenerate perturbation theory, i.e., we
solve the perturbed Hamiltonian $H={\mathcal H}+H'$ restricted to the degenerate
manifold basis $\{\phi_\mu\}$.

It is easy to see by symmetry that  $\langle k_\mu|H'|k_\mu \rangle$ is a
negative constant (independent of $\mu$)  which we call $(-\delta)$, with this
the matrix elements of the hamiltonian written in the $\{\phi_\mu\}$ basis
may be obtained

\begin{eqnarray}
\langle k_\mu|H|k_\mu \rangle &=& \langle k_\mu|{\cal H}|k_\mu \rangle+\langle k_\mu|H'|k_\mu \rangle= E_{\rm KL}-\delta ,  \nonumber \\
\langle k_\mu|H|k_{-\mu}\rangle&=& \langle k_\mu|H'|k_{-\mu}\rangle = \Delta_\|,  \nonumber\\
\langle k_\mu|H|k_{\nu} \rangle& = &\langle k_\mu|H'|k_{\nu} \rangle = \Delta_\bot , {\mathrm {if }} \,\,\nu \ne \pm \mu.
\label{eq:KLmatrix}
\end{eqnarray}
Direct calculation of the eigenvectors of the hamiltonian matrix yield a spectrum split in a singlet A$_1$ at energy $E_{\rm KL}-\delta + \Delta_\|+4 \Delta_\bot$, a triplet T$_2$ at $E_{\rm KL}-\delta  - \Delta_\|$, and a doublet E at $E_{\rm KL} - \delta  + \Delta_\|-2 \Delta_\bot$.

A possible semi-empirical solution would be to treat the three matrix
elements $\delta$, $ \Delta_\| $ and $ \Delta_\bot $ as parameters chosen to reproduce the 1s
manifold spectrum of each donor species. In Refs.~[\onlinecite{Koiller02,Baena12}]
the values $ \Delta_\|=-1.52$ meV and $\Delta_\bot=-2.16$ meV are suggested
for P donors in Si. This choice reproduces the relative energy splitting
among the different 1s manifold levels. An overall negative shift of
$\delta=4.22$ meV for P is required in order to get the actual experimental
energies. The small values of $ \Delta_\|$, $\Delta_\bot$ and $\delta$ with respect to $E_{\rm KL}$ justify the perturbative treatment described here.

The observed experimental energies can also be understood from central cell corrections, which take into account the incomplete screening of the positive
donor potential, as detailed in Sec.~\ref{cc}. The adopted central cell correction involves a single, instead of three, empirical parameter characterizing each donor species.

\subsection{Effective pseudopotential: the central cell correction \label{cc}}

Another strategy to describe the donors beyond the single valley theory is to build a pseudopotential $V_{\rm imp}(r)$ that effectively mimics the complex environment of the crystal and the screening due to the valence electrons of the donor. The suffix imp stands for the impurity type, so imp=P, As, Sb or Bi. Multivalley semiconductors are more strongly subject to this inner core potential than single valley semiconductors because the valley degree of freedom allows for the electron to concentrate closer to the nucleus. The rationale for that is simple -- each wavefunction $\phi_\mu$ spreads in k-space around ${\mathbf k_\mu}$ in a form determined by the Heisenberg uncertainty principle. The spread of the superposition of various $\phi_\mu$ in k-space tends to be larger than on the non-degenerate case, so that the real space spread may be smaller without costing any kinetic energy. Thus, multivalley electronic states tend to concentrate around the nucleus, benefiting from its potential energy and therefore lowering its total energy. A consequence is that a larger portion of the ground state energy comes from the potential very close to the nucleus, which is precisely the region where the point charge model fails.

Formally the coupling between valleys, $\langle k_\mu|H'|k_{\nu}\rangle $, could be obtained directly from the perturbation due to the donor breaking the effective mass assumptions, modeled above by the term $H'$.  The full donor  potential, $V_{\rm imp}(r)$,  has been treated in the literature within several models, a few of which are given in Fig.~\ref{fig3}.
For example, a constant replacement for the screened Coulomb potential within a central cell region of radius $R$ has been adopted in Ref.~[\onlinecite{Oliveira86}], namely, $V_{\rm imp}(r<R) = V_0$, $V_{\rm imp}(r>R) = V_H(r)$, with $R$ taken as half the nearest neighbors distance in Si and $V_0$ is a negative energy chosen to reproduce the experimental spectra.
A conceptually similar scheme, widely adopted in tight-binding (TB) calculations, is presented in Ref.~[\onlinecite{Menchero99}] for GaAs and in Refs.~[\onlinecite{Martins05,Martins02}] for Si.
It consists of taking the usual screened Coulomb potential $\sim 1/r_i$ to correct the on-site energies at atomic positions ${\mathbf r}_i\ne 0$. For the impurity site the Coulomb correction would diverge, which is incompatible with TB, so a finite correction $U_0$ is chosen to adjust the correct binding energies of each donor.

Realistic descriptions of the donor Coulomb potential should be consistent with the expected  limits: $V_{\rm imp}(r\to \infty)=-e^2/4\pi\epsilon_{\rm Si}r = V_H(r)$ and $V_{\rm imp}(r\to 0) =  -e^2/4\pi\epsilon_0 r$. A convenient interpolation~[\onlinecite{Wang73}] is adopted here, see Fig.~\ref{fig3}, namely
\begin{eqnarray}
  V_{\rm imp}(r)& =& -{\frac{e^2}{4\pi r}}\left[{\frac{1}{\epsilon_{\rm Si}}} + \left({\frac{1}{\epsilon_0}}-{\frac{1}{\epsilon_{\rm Si}}}\right)e^{-r/r_{cc}}\right]  \nonumber \\
  &=& V_H (r) + V_{cc}(r) \label{eqrcc}
\end{eqnarray}
This involves a single donor-dependent parameter, $r_{cc}$, giving the range around the donor site where the central cell correction is effective, as opposed to the three quantities required for the description given in  Eq.~(\ref{eq:KLmatrix}). Values for $r_{cc}$ appropriate for P, As, Sb and Bi donors in Si are given in Table~\ref{radii}. Note that while Bohr radii are of the order of a few nm, $r_{cc}$ is a factor of 10 smaller. Therefore the deviation of the plain Coulomb potential from the central-cell corrected one is restricted to a small region around the donor, as illustrated in Fig.~\ref{fig2}(b).

\begin{table}[h!]
\centering
\begin{tabular} {c  c  c  c  c  c}
\hline \hline
Donor        &$E^0_{\rm exp}$(meV)~[\onlinecite{Grimmeiss82}] & $E^0_{\rm calc} $(meV)  &   $r_{cc}$ (nm) & $a_{cc}$ (nm)              \\ \hline
                 & -45.58 (A$_1$)  &  $ E^0_{\rm exp}$  &                                  &   \\
P$^0$       & -33.90 (T$_2$)     &-33.12                                       &    0.115                 & 1.106  \\
                 & -32.60 (E)           & -33.12                                  &                            &   \\
                 &                            &                                                   &                            &                             \\
                 & -53.77 (A$_1$)  &     $ E^0_{\rm exp}$  &                                       & \\
As$^0$    & -32.65 (T$_2$)     & -32.15                                         &    0.126                   & 0.815   \\
                 & -31.40 (E)         & -32.15                                     &                             &    \\
                 &                           &                                                  &                             &                               \\
                 & -42.71 (A$_1$) &      $ E^0_{\rm exp}$                             &                             &     \\
Sb$^0$    & -33.00 (T$_2$)     &-32.04                                            &    0.108                  & 1.241   \\
                 & -30.60 (E)        & -32.04                                       &                            &    \\
                 &                            &                                                   &                            &                             \\
                       & -71.00 (A$_1$) &      $ E^0_{\rm exp}$                             &                             &     \\
Bi$^0$    & -31.92 (T$_2$)     &N/A                                         &    0.144                  & 0.58   \\
                 & N/A (E)        & N/A                                      &                            &    \\

\hline \hline
\end{tabular}
\caption{Experimentally measured~[\onlinecite{Grimmeiss82}]  values $E^0_{\rm exp}$ of the 1s manifold energy states for the neutral isolated donor are compared to the calculated values $E^0_{\rm calc}$ obtained from the central cell corrected potential Eq.~(\ref{eqrcc}). The energy of the central cell corrected state A$_1$ is fitted to the corresponding experimental value by choosing an appropriate $r_{cc}$. The E and T$_2$ levels are assumed degenerate.
The calculation of the levels E and T$_2$  involves $\tilde E$, the center of gravity of the sixfold 1s manifold, which is taken from experiment (see text), but is not available for Bi. Note that the experimental and calculated values for the T$_2$ and E energies do not depend strongly on the donor species and are very similar to the single valley result $E_{\rm KL}$, in contrast to the value of the ground state A$_1$ energy. As far as we know, the Bi E level is not available from  experiment, but the T$_2$ level is consistent with these observations. For each donor, the effective Bohr radius corrected by the central cell $a_{cc}$ is also given.}
\label{radii}
\end{table}

\section{Neutral Donor -- D$^0$ center \label{dzero}}

The coupling matrix elements calculation may be simplified by assuming spherically symmetric envelopes. It is possible to conciliate the simpler hydrogenic description in Eq.~(\ref{eq-psih}) with the ground state energy $E_{\rm KL}$ obtained from Eq.~(\ref{eq-psikl})  by choosing a value of the effective mass $m^*_{\rm Si} =
0.3~m_e$. Note that $m^*_{\rm Si}$ is very
close to the geometric mean  $(m_L \times m^2_T)^{1/3}$ = 0.32. This
approximation results, for the uncoupled valleys, into a sixfold degenerate
ground state with ground state energy $E_{\rm KL}$. 

Now we take into account the multivalley structure of the Si conduction band and the central cell correction described in Eq.~(\ref{eqrcc}). The only parameter we are left with is $r_{cc}$, the central cell potential characteristic length, chosen such that the ground state A$_1$ experimental energy is reproduced for each of the donor species, as given in Table~\ref{radii}.

The T$_2$ and E states are much less sensitive to the effect of $V_{cc}$ since both have nodes at the impurity site. This is confirmed by their approximate degeneracy and energy very close to $E_{\rm KL}$ (see Table~\ref{radii}).  
We take the approximation of exact fivefold degeneracy of these levels, obtained assuming $ \Delta_\| = \Delta_\bot = \Delta$, so that E(A$_1$) = $E_{\rm KL}-\delta +5 \Delta$ and E(T$_2$,E) = $E_{\rm KL}-\delta + \Delta$. The mean value of the six levels is taken from experiments and identified with $\tilde E = E_{\rm KL}-\delta$  which leads to $\delta = E_{\rm KL}-\tilde E$. It then follows that $\Delta =$[E$_{\rm KL} - $E(A$_1)]/5 $ allowing to determine the excited states energies.

Combining spherically symmetric envelopes with the Bloch functions obtained from the expansion coefficients given in Ref.~[\onlinecite{Saraiva11}] yield the electronic densities for the A$_1$ and T$_2^x$ bound donor states of P in Si shown in Fig.~\ref{fig2}(c). Note that the charge distribution of the T$_2^x$ state has a preferential alignment along the $x$ direction and an
exact node at the origin (\textit{idem} to E states). Due to this node, any contact interaction (interaction potentials that are only finite at the immediate vicinity of the donor impurity) will have a reduced effect on the T$_2$ and E states -- this justifies why these states are not strongly affected by the central cell correction as well as indicates that low hyperfine coupling is to be expected in these states. For this reason, the T$_2$ and E states energies are very similar to $E_{\rm KL}$. The results are summarized in Table~\ref{radii}.

\section{Two Electrons in One Donor -- D$^-$-center\label{dminus}}
In analogy with the hydrogen  atom,  singly ionized negative (H$^-$)  or positive (H$^+$) states are also of practical and theoretical interest. The negative ion H$^-$ binding energy is defined as the energy required to remove one electron from the ion $E_B^{H^-}=E_{H^0}-E_{H^-}$. For hydrogen $E_B^{H^-}=0.055 E_H$.
The analogous D$^-$ center is specially important in Si nanoelectronics because its binding energy is very small and can be strongly affected by gates~[\onlinecite{ourD-}]. The presence of this bound state has been identified in transport experiments in single-atom transistors~[\onlinecite{sellierPRL2006}].

We may also regard D$^-$ centers as He-like systems~[\onlinecite{bethe-salpeter-book}], the main technical difficulty to obtain the spectrum consists in the evaluation of electron-electron repulsion terms. Variational wavefunctions with a large number of parameters, as used by Hylleraas~[\onlinecite{Hylleraas}] and others, are able to reproduce the experimental value of $E_B^{H^-}$.

In general, the ground state wavefunction is composed of an inner orbital with a Bohr radius similar to the electron bound to a neutral donor, and an outer orbital with a much larger radius, due to the screening of the Coulomb potential produced by the occupied inner orbital.

 Within a single valley and isotropic mass approach for negatively ionized centers in Si, we can just rescale the exact results for $H^-$: $E_B^{D^-}=0.055 E^* \sim 1.7 $ meV, which gives a good estimate for Phosphorus and Arsenic centers: $E_B^{P^-}=1.7$ meV, $E_B^{As^-}=2.05$ meV as measured in photoconductivity experiments~[\onlinecite{narita1982}].

The inclusion of the mass anisotropy and the valley degeneracy in Si would increase the binding energy of the $D^-$ centers even if valley-orbit interactions are not considered~[\onlinecite{kamimura1979}]. If no central cell corrections are included, each electron remains  in a different valley. The lowest energy configuration is attained when the two electrons are located in perpendicular valleys  (as m$_\parallel$ and m$_\perp$ are rotated in perpendicular valleys). Mass anisotropy then implies that the envelopes corresponding to the two orbitals do not overlap as much as they would for isotropic wavefunctions reducing the electron-electron repulsive interaction.

Valley-orbit interactions and central-cell corrections have been treated in the literature for the D$^-$ problem under different approximations~[\onlinecite{Oliveira86,larsen1981,inoue2008}].
Larsen~[\onlinecite{larsen1981}] assumed that only the inner orbital electron is subject  to valley-orbit coupling,  behaving like a neutral donor electron. The isotropic condition  $\Delta_{\parallel} = \Delta_{\perp}$ was also used. The effect of the valley-orbit coupling on the inner electron is to spread it in a symmetric ($A_1$) combination of valleys. If both electrons have the same valley composition, the binding energy decreases with respect to them being in perpendicular valleys. Oliveira and Falicov~[\onlinecite{Oliveira86}] went a step further by including the full 1s multiplet in the description of the bound electrons and a constant valley-orbit coupling different from zero only in a central-cell region. More recent calculations~[\onlinecite{inoue2008}] included the valley degeneracy but not the valley-orbit splitting.

Here we treat the $D^{-}$ donors on the same level of approximation as the neutral donors, hence neglecting the mass anisotropy. We take the simplest possible form for the spin singlet ground state trial function (normalized):
\begin{equation}
\psi_{D^-}={1 \over {\pi \sqrt{2 a_1^{3} a_2^{3}}}}\left(e^{- r_1/a_1}e^{-r_2/a_2}+e^{-r_2/a_1}e^{-r_1/a_2}\right)\, ,
\label{eq:Dminuswf}
\end{equation}
and consider the central-cell corrected potential $V_{\rm imp}(r)$, as defined in Eq.~(\ref{eqrcc}). The correction term $V_{cc}$ is completely determined by the value of $r_{cc}$ chosen  from the corresponding neutral donor data (see Table~\ref{radii}). Both $a_1$ and $a_2$ are calculated variationally to minimize the expectation value of the energy for the two-electrons hamiltonian: $H_{D^-}=K_1+K_2+V_{\rm imp}(r_1)+V_{\rm imp}(r_2)+e^2/\epsilon_{\rm Si} r_{12}$ where $K_i$ is the kinetic energy of electron $i$  and $V_{\rm imp}(r_i)$ is the central cell corrected potential. The last term is the electron-electron repulsion. The energies are calculated following the prescription for He atoms given in Ref.~[\onlinecite{bethe-salpeter-book}].

If central cell corrections were ignored, the variational wavefunction in Eq.~(\ref{eq:Dminuswf}) would lead to a binding energy $E_B^{D^-}=0.027 E^*=0.8$ meV, see Ref.~[\onlinecite{ourD-}]. The inclusion of the central cell correction allows for a significant improvement of the binding energy,  see Table~\ref{table:Dminus}. The Table also shows the value of the charging energy $U=E_{H^-} - 2E_{H^0}$ which can be measured in transport spectroscopy experiments in single atom transistors~[\onlinecite{ourD-}].

\begin{table}[h!]
\centering
\begin{tabular} {c c c  c  c  c  c  c  c}
\hline \hline
Donor        &  $E^-_{\rm calc}$ & $E^-_{\rm exp}$ & $E_{B\rm (calc)}$ &  $E_{B\rm (exp)}$ &$ U_{\rm calc}$  &$U_{\rm exp}$ & $a_1$  & $a_2$          \\ \hline

P$^-$       & -47.10     &-47.29              & 1.526  & 1.7   &     43.86     &     43.00         &    1.041                 & 4.851  \\

                 &                            &                 &   &  &                 &                 &                            &                             \\

As$^-$    & -56.44     & -55.81             & 2.69 & 2.05 &       51.06     &    51.71           &    0.737                   & 3.516   \\

                 &                           &                & & &                     &           &                             &                               \\

Sb$^-$    & -44.00     & N/A               & 1.29 & N/A &     41.44      &    N/A          &    1.177                 & 5.349   \\

                 &                           &                & & &                     &           &                             &                               \\

Bi$^-$    & -78.55     & N/A               & 7.55 & N/A &     63.45      &    N/A          &    0.48                 & 2.16   \\

\hline \hline
\end{tabular}
\caption{Calculated and experimental~[\onlinecite{narita1982}]  energies and variational parameters for the negatively charged donors. All energies are given in meV and lengths in nm. $E_B$ refers to the first ionization energy, namely, the energy required to ionize the less bound electron $E_B=E^0-E^-$. The charging energy $U$ is defined as $U=E^--2E^0$. The parameters $a_1$ and $a_2$ are the Bohr radii obtained variationally. The radius $a_1$ corresponds to the inner orbital, and is very similar to $a_{cc}$ in Table~\ref{radii}. The parameter $a_2$ is the radius of the outer orbital which is much less bound than the inner one. To the best of our knowledge, the experimental values for Sb and Bi are not known.}
\label{table:Dminus}
\end{table}

Note the very good agreement between calculated and experimental (when available) values  in Table~\ref{table:Dminus}, indicating the quantitative validity and transferability of the central cell correction given here and validating our predictions for Sb and Bi.

\section{One Electron and Two Donors -- D$_2^+$ center\label{d2plus}}

The problem of two donors sharing a single electron is analogous to the problem of an ionized $H_2^+$ molecule in vacuum. If it was identical it would be possible to solve this problem exactly through the transformation into spheroidal coordinates~[\onlinecite{slater-book}]. However, as discussed for single donors, the effective mass anisotropy and the valley orbit coupling in silicon complicate the spectrum. Many accounts of these effects are available in the literature. Early approaches employed an altered form of the spheroidal coordinates to accomodate mass anisotropy to some extent~[\onlinecite{miller1960}].

We steer the analogy with the $H_2^+$ molecule along another direction, namely  writing the eigenstates of this problem as bonding and anti-bonding molecular orbitals based on the single donor (atomic) problem, i.e., taking molecular wavefunctions as linear combination of atomic orbitals (LCAO). The two donors, referred to as $A$ and $B$, located at positions $\mathbf{r_A}$ and $\mathbf{r_B}$ are separated by a vector $\mathbf{R}=\mathbf{r_B}-\mathbf{r_A}$.
An obvious difference between the substitutional donor pair and the molecular problems is the meaning and range of possible values of $\mathbf R$. For donors, $\mathbf{R}$ is fixed at the sample fabrication and doping  stage and coincides with a lattice vector, while in the free molecule situation, $R=|\mathbf{ R}|$ is given by the minimization of the molecule energy.

The effective mass hamiltonian of a donor pair reads
\begin{equation}
H_{DD} ({\mathbf r})=-\frac{\hbar^2 \nabla^2}{2m^*}+V_{\rm imp}(r_A)+V_{\rm imp}(r_B).
\label{eq:h_1_e_2_d}
\end{equation}
The matrix elements of this hamiltonian are calculated in the LCAO basis, \textit{i.e.}, in the basis set defined by the single donor wavefunctions \{A$_1$,T$_2$, E\} given in Eq.~(\ref{eq:single-donor-wf}), centered at $\mathbf{r_A}$ and $\mathbf{r_B}$. We expect this basis to give a fair account of the lowest electronic states.
The hamiltonian is hence a 12$\times$12 matrix, which for convenience we break into four $6 \times 6$ blocks,
\begin{equation}
H=
\left[{\begin{array}{cc}
 H_{AA} & H_{AB}\\
 H_{BA} & H_{BB}
\end{array}}\right]
\label{eq:matrix}
\end{equation}
Blocks $H_{AA}$ and $H_{BB}$ are not strictly diagonal since the problem of two donors has lower symmetry than a single donor. But a direct calculation shows that the off-diagonal terms in these 2 blocks are  vanishingly small compared to other non-zero terms for interdonor distances larger than the lattice constant $a_{\rm Si}$. We therefore take the diagonal blocks to be diagonal, which shows that  the donor eigenstates basis set is more convenient than the $\{\phi_\mu\}$ for the pair treatment. These blocks, therefore, represent the on-site energies while the tunnel coupling terms contribute to the blocks $H_{AB}$ and $H_{BA}$. The diagonal blocks $[H_{AA}]=[H_{BB}] $ considering a basis ordered as in the sequence in Eq.~(\ref{eq:single-donor-wf}):
\begin{equation}
[H_{ii}]=\left(\begin{array}{cccccc}
\varepsilon^{onsite}_{CC} & 0 & 0 & 0 & 0 & 0 \\
0& \varepsilon^{onsite}_{sv} & 0 & 0 & 0 & 0  \\
0&0& \varepsilon^{onsite}_{sv} & 0 & 0 & 0 \\
0& 0& 0 & \varepsilon^{onsite}_{sv} & 0 & 0 \\
0 & 0 & 0 & 0 & \varepsilon^{onsite}_{sv} & 0\\
0 & 0 & 0 & 0 & 0 & \varepsilon^{onsite}_{sv}
\end{array} \right).
\end{equation}
The onsite energy for the A$_1$ state $\varepsilon^{onsite}_{CC}$ is the ground state energy for the neutral donor including the central cell contribution from the $A$ donor, corrected by a long range classical term $V^\prime_{\rm imp}=\langle \Psi_{A_1}(r_A)| V_H(r_B) + V_{CC}(r_B)|\Psi_{A_1}(r_A)\rangle$ to take into account the $B$ donor potential. For T$_2$ and E, $\varepsilon^{onsite}_{sv}$ is the single valley $E_{\rm KL}$ corrected by  $V^\prime_{H}=\langle \Psi_{T / E}(r_A)| V_H(r_B) |\Psi_{T / E}(r_A)\rangle$. Following the argument in Sec.~\ref{dzero}, the central-cell correction on T$_2$ and E is neglected because these states have nodes at all Si sites.

The off-diagonal blocks are related by the hermiticity condition $H_{AB}=H_{BA}^\dagger$. Each term is a summation over integrals of the type
\begin{widetext}
\begin{equation}
\langle \phi_\mu ({\mathbf r_A})|H_{DD}|\phi_\nu ({\mathbf r_B})\rangle = \int  F(r_A) e^{- i {\mathbf k_\mu . r_A}} u^*_\mu({\mathbf r_A}) H_{DD} F(r_B) e^{ i {\mathbf k_\nu . r_B}} u_\nu ({\mathbf r_B}) \mathrm{d}^3 {\mathbf r}.
\label{eq:tunnel}
\end{equation}
\end{widetext}

These matrix elements can be straightforwardly calculated using the plane wave expansion of the Bloch functions (see Ref.~[\onlinecite{Saraiva11}]), and no further approximation is needed. Nevertheless, it is useful to simplify these integrals  through some well tested approximations: (i) neglecting the matrix elements for $\mu\ne\nu$ , since rapidly oscillatory integrands are involved; (ii) taking $u^*_\mu u_\mu \approx 1$, as suggested in Ref.~[\onlinecite{Saraiva11}]. Under these assumptions we obtain
\begin{equation}
\langle \tilde \phi^{A,i}_\mu | H_{DD} ({\mathbf{r}}) | \tilde\phi^{B,j}_\nu\rangle = \delta_{\mu,\nu} e^{i \mathbf{k_\mu}\cdot \mathbf{R}} t_{sv}(R),
\label{eq:tunnel_simple}
\end{equation}
where $t_{sv}$ is the single valley tunnel coupling
\begin{equation}
t_{sv}(R) = \int  F(r_A) H_{DD} F(r_B) {\mathrm d}^3 {\mathbf r}.
\label{eq:tunnel_sv}
\end{equation}

\begin{figure}
\leavevmode
\includegraphics[clip,width=0.48\textwidth]{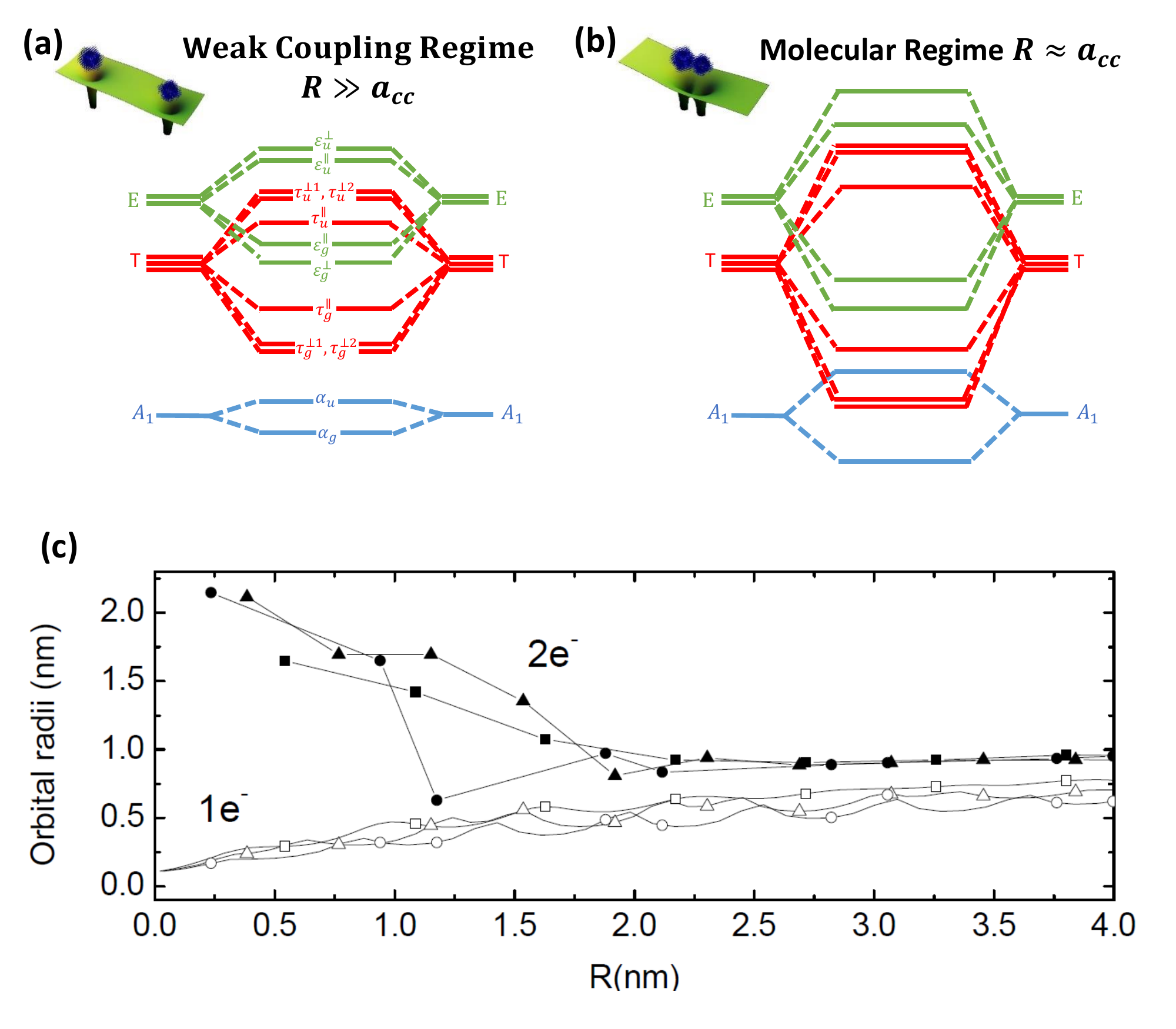}
\caption{Schematic behavior of the molecular orbitals spectra from the 1s manifold with different valley compositions in the (a) weak coupling regime for a single electron and (b) molecular (strong coupling) regime for a single electron. (c) Variational radii calculated to minimize the energy of the donor molecule oriented along the $\langle 100 \rangle$, $\langle 110 \rangle$, and $\langle 111 \rangle$ directions (squares, triangles and circles respectively). The variational radii for the ionized $a_{D_2^+}$ (empty symbols) and neutral $a_{D_2^0}$ (solid symbols) molecule are shown.}
\label{fig4}
\end{figure}

The effect of the hopping blocks $H_{AB}$ is analogous to the H$_2$ molecule -- at distances much larger than $a_{cc}$, this block is negligible and the states centered around sites A and B are degenerate, while for distances comparable to the one-atom wavefunction extension, {\textit gerade} and {\textit ungerade} combinations of the localized orbitals form, leading to split energies, see Fig.~\ref{fig4}(a) and (b). At smaller interdonor distances, the \textit{gerade - ungerade} splitting energy becomes comparable to the A$_1$-T valley-orbit splitting and a level inversion of the first excited state is obtained. This result is discussed theoretically in Ref.~[\onlinecite{klymenko2014}] and confirmed experimentally in Ref.~[\onlinecite{dehollain2014}].

Unlike the H$_2$ molecule problem, it is possible for the antisymmetric state (referred to as antibonding in the context of H$_2$ in vacuum) to be the ground state here, since the hoppings are not necessarily real negative numbers due to the oscillatory phase $\exp(i \mathbf{k_\mu}\cdot\mathbf{R})$. Still, we refer to the lower molecular orbital as {\textit gerade} and the higher as {\textit ungerade}.

The same oscillatory phase may also lift the degeneracies of the T$_2$ and E states. For instance, if the pair alignment is along the $x$ direction, the states $T_2^y$ and $T_2^z$ are still equivalent, while the state $T_2^x$ will have a symmetric-antisymmetric splitting that oscillates as a function of $R$. This effect is further enhanced in the presence of the effective mass anisotropy, as seen in Ref.~[\onlinecite{klymenko2014}]. The effective mass anisotropy also impacts the interdonor distance at which the valley inversion of the first excited state occurs.

It is known from the H$_2$ problem that the molecular orbital approximation gives accurate results only if the variational wavefunction radius is taken to minimize the expectation value of the complete hamiltonian containing the two protons. We do the same here, obtaining a variational radius $a_{D_2^+}(\mathbf{R})$, which converges to the single atom orbit at large distances $a_{D_2^+}(R\to\infty)=a_{cc}$.

\begin{figure}
\leavevmode
\includegraphics[clip,width=0.48\textwidth]{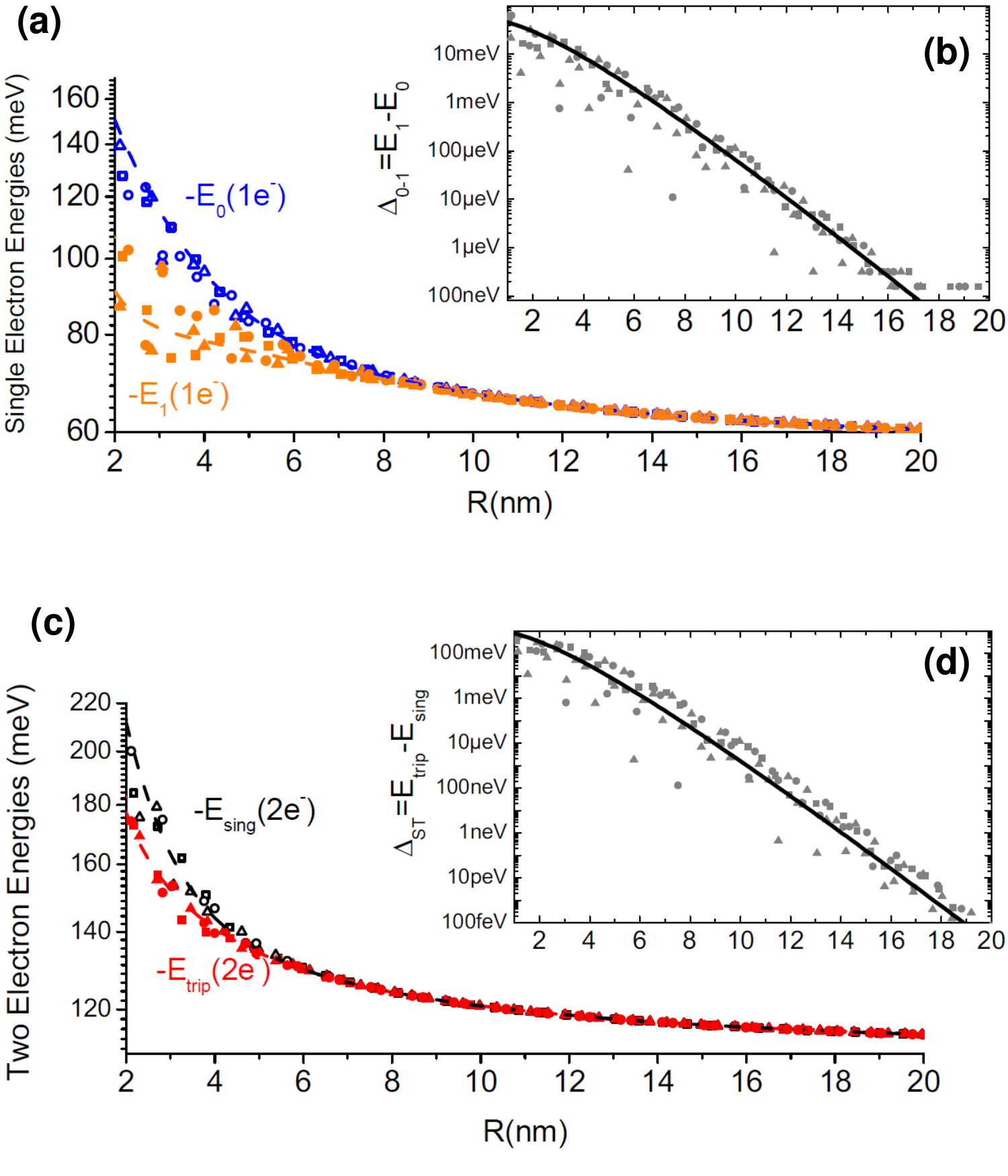}
\caption{Ground and first excited state energies for one and two electrons. (a) The ground state is the bonding $\alpha_g$ orbital and the first excited is the antibonding $\alpha_u$ in the weak coupling regime ($R\gg a_{cc}$) or the bonding $\tau_g$ in the strongly interacting regime ($R\ll a_{cc}$), see Fig.~\ref{fig4}. The dashed lines are the LCAO results using single-valley central cell corrected 1s wavefunctions, and taking the H\"uckel approximation for the off-diagonal integrals. (b) The energy separation between ground and first excited states $\Delta_{0-1}$, which coincides with the tunnel coupling in the weak coupling regime. Squares, triangles and circles correspond to the $\langle 100 \rangle$, $\langle 110 \rangle$ and $\langle 111 \rangle$ molecule orientations respectively. (c) Two electron energies. The ground state is a spin singlet composed by two electrons in the bonding $\alpha_g$ orbital. The first excited state is a spin triplet, with one electron in $\alpha_g$ and the second electron in either $\alpha_u$ or $\tau_g$ (in the weakly or strongly interacting regimes, respectively). The dashed lines show the Hubbard approximation. (d) The singlet-triplet separation $\Delta_{ST}$, which coincides with the spin-spin Heisenberg exchange coupling in the case of two electrons.}
\label{fig5}
\end{figure}

The resulting energies for the ground [$E_0(1e^-)$] and the first excited [$E_1(1e^-)$] states are plotted with symbols in Fig.~\ref{fig5}(a).  The squares, triangles and circles correspond to the three different molecule orientations considered $\langle 100 \rangle$, $\langle 110 \rangle$ and $\langle 111 \rangle$. The oscillations with $R$ are due to the intervalley interference which produce the incommensurate oscillations on the single donor wavefunctions, illustrated in Fig.~\ref{fig2}(c). This leads to subtle oscillations in the total energy, but significant oscillations in the energy difference $\Delta_{0-1}=E_1(1e^-) - E_0(1e^-)$ [Fig.~\ref{fig5}(b)].

\section{Two Electrons in Two Donors -- D$_2^0$ center }
\label{d2}

The quantum mechanical solution to the problem of two impurities with two interacting electrons is significantly less studied, even though it is the most commonly found configuration since it is neutral. The electron-electron correlations are hard to describe and add significant complexity to the picture described in the previous section. The most systematic way to calculate with high accuracy the two electron energies is to build a full Configuration Interaction (CI) wavefunction from the molecular orbitals discussed for the D$_2^+$ problem.

Now the suitability of truncating our basis set at the 1s manifold becomes clear. This manifold contains 6 atomic orbitals (one for each valley configuration), leading to 12 molecular orbitals and 144 two-electron Slater determinants. This way, the CI matrix for this problem would contain 20736 elements, each consisting of a few Coulomb integrals -- six-dimensional improper numerical integrals with an integrand diverging at all points where the electron-electron distance is null $r_{12}=0$. Even a small increment of the atomic basis set increases the numerical demand of this problem significantly.

Still the problem stated above is too hard to be solved without any algebraic maneuvers and approximations. The first step is to identify elements that are null by symmetry. In the case of two electrons, one can always write down the wavefunction as a product of orbital and spinorial parts  (with three or more particles these properties are necessarily entangled). Since the total spin operator and its projection both commute with the hamiltonian, it is wise to rewrite the molecular orbital basis set to be composed of eigenstates of these operators (singlets and triplets). This constitutes the so-called SACI  (Spin Adapted Configuration Interaction)~[\onlinecite{wang2009}], which leads to a CI matrix composed by one $78\times78$ singlet block and one $66\times66$ triplet block, with null off-diagonal blocks. This cuts the computational effort by roughly a factor of two.

These dense singlet and triplet matrices are still too challenging if all the Coulomb integrals are calculated numerically. A very robust quantum chemistry approximation to overcome this problem is to expand the Slater-type orbitals into a series of $N$ gaussians with given radii. This approach converges quickly with $N$, and for the calculations performed here for interdonor distances up to $R=20$ nm, the energies were well converged with $N=3$. For larger distances the number of gaussians increases due to the ill described tail decay~[\onlinecite{saraiva07}]. The results for the numerical calculation of the singlet and triplet energies are presented in Fig.~\ref{fig5}(c) while the Bohr radii $a_{D_2^0}$ resulting from the variational minimization are in Fig.~\ref{fig4}(c). The singlet triplet separation $\Delta_{ST}$, in Fig.~\ref{fig5}(d) reveals the same oscillations due to valley interference~[\onlinecite{wellard2005,Koiller02}] as the single electron excitation energy $\Delta_{0-1}$.

The CI method is considered the standard model of modern quantum chemistry. In principle, arbitrarily accurate results may be obtained if a large enough single particle basis set is adopted. In practice, though, a truncated basis set could lead to largely incorrect results -- specially if the chosen basis set is inadequate. For instance, in Sec.~\ref{dminus} we adopted a simple two electron wavefunction, Eq.~(\ref{eq:Dminuswf}), which describes with a single orbital the energy of the D$^-$ center with great accuracy. If instead a regular CI basis set is constructed from the multiorbital LCAO method, the variational energy obtained is much higher and the result is qualitatively wrong -- an unbound D$^-$ state is obtained, in contrast to the experimentally measured bound D$^-$. 

\section{Discussion }
\label{sec:discussion}
\subsection{Simplified Models and effective Hamiltonians}
\label{simple}

In certain studies, the description of donor states in Si  based on microscopic model hamiltonians as presented in the previous sections may not be necessary or useful. Instead, simple models may be more adequate. In this section we investigate possibilities to map the problem into simplified models.

\subsubsection{D$_2^+$: Single Orbital LCAO and the H\"uckel Approximation.}

Inclusion of higher excited orbitals improves the accuracy of the calculations, but in the simplest picture a single 1s orbital is considered, so that only two molecular-orbitals are obtained. Another simplifying assumption is to disregard the effect of valley physics. This approximation disregards the oscillations in the tunnel coupling. Instead, the tunnel coupling becomes a real monotonic function of the interdonor distance, and the molecular states are the symmetric and antisymmetric combinations of the atomic orbitals
\begin{eqnarray}
|\alpha_g\rangle=\left(|A_1^A\rangle+|A_1^B\rangle\right)/\sqrt{2},\\
|\alpha_u\rangle=\left(|A_1^A\rangle-|A_1^B\rangle\right)/\sqrt{2}.
\end{eqnarray}
The energies of these two states are $E=E_{onsite}\pm |t|$, where $E_{onsite}$ is the onsite energy of the atomic orbital and $t$ is the tunnel coupling, or hopping parameter.

The onsite energy is readily obtained. We may break down the hamiltonian as $H=H_A+V_B$, so that $H_A$ is identical to the single donor case, and $\langle A_1^A|H_A|A_1^A\rangle=E_0$ is simply the single electron in a single donor energy, known with great accuracy from experiments, hence we take $E_0=E_{exp}$. The second term $\langle A_1^A|V_B|A_1^A\rangle$ could in principle be calculated exactly, but since our wavefunction description is not particularly accurate, specially for $R\approx a_{cc}$, we take a simplified point charge interaction, such that $\langle A_1^A|V_B|A_1^A\rangle=-e^2/4\pi\epsilon_{\rm Si} R$. $e^2/4\pi\epsilon_{\rm Si} \approx 133.5$~meV~nm.

The tunnel coupling could also be explicitly calculated, but a simpler result is obtained using the H\"uckel approximation. The argument behind this approximation is that the atomic orbital is an exact eigenstate of the single donor part $H_A$, and the second center potential $V_B$ may be treated approximately as a percentual correction over the $H_A$. In our case, we go further in the approximation and take $\langle A_1^B|V_B|A_1^A\rangle=0$, leading to $t=\langle A_1^B|H|A_1^A\rangle\approx\langle A_1^B|H_A|A_1^A\rangle=E_0 S(R,a_{cc})$, where the overlap function is $S(R,a_{cc})= \langle A_1^B|A_1^A\rangle=e^{-R/a_{cc}}(1+R/a_{cc}+R^2/3a_{cc}^2)$.

A comparison between these approximations and the results for the complete numerical calculation, shown in Fig.~\ref{fig5}(a) and (b), reveals that this approximation is well suited for estimating the order of magnitude of the tunnel coupling (estimated as the separation between the ground and first excited states $\Delta_{01}$ in our complete multiorbital LCAO method of Sec.~\ref{d2plus}), as well as the ground and first excited energies to an accuracy of approximately 10-15\%, as seen in Fig.~\ref{fig5}. The spread of the tunnel coupling in the numerical results is a product of valley interference, which is disregarded here.

\subsubsection{D$_2^0$: Hubbard Model.}

The energy for two electrons may be calculated with similar arguments. The electron-electron repulsion supresses the probability of double occupation of the same donor. We take into account the correction to the singlet ground state due to the virtual occupation of this excited state, but discard any contributions from the triplet with both electrons at the same site since it has a much higher onsite energy. This is one of the ingredients of the so-called Hubbard approximation.

This way, the triplet energy is the easiest to calculate, consisting only of the binding energies of the two electrons to each respective donor and the classical attraction/repulsion across donors/electrons. Hence, $E_{trip}(2e^-)=\varepsilon_{(1,1)}= 2E_0-e^2/4\pi\epsilon_{\rm Si} R$.

The singlets are obtained diagonalizing the hamiltonian written in the basis $\{|s^A({\mathbf r_1})s^A({\mathbf r_2})\rangle, (|s^A({\mathbf r_1})s^B({\mathbf r_2})\rangle+|s^B({\mathbf r_1})s^A({\mathbf r_2})\rangle)/\sqrt{2}, |s^B({\mathbf r_1})s^B({\mathbf r_2})\rangle\}$, which reads
\begin{equation}
[H]_{singlets}=\left(\begin{array}{ccc}
\varepsilon_{(2,0)}+U & \sqrt{2}t & 0\\
\sqrt{2}t& \varepsilon_{(1,1)} & \sqrt{2}t  \\
0&\sqrt{2}t&\varepsilon_{(0,2)}+U
\end{array} \right).
\end{equation}
We take the coupling between the (1,1) and (2,0)/(0,2) states to be determined by the single particle hopping $t$, which is calculated within the H\"uckel approximation discussed previously. This approximation is not necessarily accurate, since the size of the wavefunction of (2,0)/(0,2) is most likely comparable to the size of the doubly occupied single donor ion $D^-$. The most important parameter for the Hubbard model is the onsite charging energy $U$, which we take to be the experimental charging energy of the D$^-$ single dopant discussed in Section~\ref{dminus} (with exception of Sb and Bi, for which we adopt the results calculated in Sec.~\ref{dminus}).

By symmetry we have $\varepsilon_{(2,0)}=\varepsilon_{(0,2)}$ in the absence of detuning fields. The electrostatic arrangement of the (1,1) and the (2,0)/(0,2) charge configurations is significantly different, and therefore the onsite energy of these two states is not the same. On the other hand, the large same-site charging energy $U$ leads to strong Coulomb blockade, and the occupation of the (2,0) and (0,2) states is only virtual. We argue that the error imposed by the approximation $\varepsilon_{(1,1)}=\varepsilon_{(2,0)}$ has a very small impact in the ground state energy [composed almost exclusively by the (1,1) configuration] as well as the singlet-triplet separation $\Delta_{ST}$. These assumptions are confirmed \textit{a posteriori}, see the comparison between the numerical and the approximated values in Fig.~\ref{fig5}(c) and~(d).

\begin{table}[h!]
\centering
\begin{tabular} {r  c  l }
$E(1e^-)$ & = & $E_0 - \frac{133.5}{R}-t(R,a_{cc})$\\ \\
$E(2e^-)$ & = & $2E_0 - \frac{133.5}{R} - J(R,a_{cc})$ \\ \\
$t(R,a_{cc})$ & = & $E_0 e^{-R/a_{cc}} (1+\frac{R}{a_{cc}}+\frac{R^2}{3a_{cc}^2})$ \\ \\
$J(R,a_{cc})$ & = & $2 \frac{t(R,a_{cc})^2}{U}$\\ \\
\end{tabular}
\begin{tabular} {c  c  c  c }
\hline \hline
$$
Donor  & $E_0$(meV) &  $a_{cc}$(nm) & $U$(meV)   \\ \hline
P          & -45.58         &  1.11           &  43.0           \\
As        & -53.77         &  0.82           &  51.7           \\
Sb        & -42.77         &  1.24           &  41.4           \\
Bi         &  -71.00        &  0.58           &  63.45           \\

\hline \hline
\end{tabular}
\caption{Explicit expressions (see Sec.~\ref{simple}) for the energies of the ionized donor molecule $E(1e^-)$ and the neutral donor molecule $E(2e^-)$ as a function of the distance $R$ between the donors and the Bohr radius calculated for the neutral donors $a_{cc}$. Input parameters for the numerical estimates are taken from Tables~\ref{radii} and~\ref{table:Dminus}. The quantity $133.5$ is given in units of meV  nm (see text).}
\label{tab:simple}
\end{table}

\subsection{Implications for experiments \label{expt}}

These results indicate possible directions for the design and characterization of one and two donor quantum devices. They may also bring new insights in specific aspects of larger systems such as donor clusters~[\onlinecite{weber2014}], impurity chains~[\onlinecite{prati2012}] and $\delta$-doped systems~[\onlinecite{shamin2014}]. Recent experimental results confirm the adequacy and validity of the present theory~[\onlinecite{dehollain2014,Gonzalez-Zalba2014}].

\subsubsection{Transport through coupled donors.}

One of the most reliable techniques for probing single or few impurities is quantum transport spectroscopy~[\onlinecite{sellierPRL2006,Gonzalez-Zalba2014,Gonzalez-Zalba2012,lansbergen2008,pierre2009,roche2013}]. From these measurements the energy differences between occupation numbers (the chemical potential) are accessed directly.

From the expressions in Table~\ref{tab:simple}, one can see that the energy of one and two electron states contain the same long distance electrostatic term. Therefore, the difference between these two energies -- i.e. the first ionization energy -- indicates there is quantum interaction between the two donors. If the tunnel coupling between the donors is not comparable to the binding energy, then the first ionization energy is essentially the same one would obtain for a single donor, i.e. $E(2e^-)-E(1e^-)\approx E_0$. We refer to this as the \emph{weak coupling regime}~[\onlinecite{roche2013}]. This is the most convenient regime for electrical control of the charge distribution and spin-spin interactions necessary for quantum computation with electron spins.

The opposite regime consists of small distances -- referred to as the \emph{molecular regime}~[\onlinecite{Gonzalez-Zalba2014}] -- in which an electronic cloud is shared between the sites $A$ and $B$ with no classically forbidden region between the two donors. The first ionization energy departs strongly from the single donor binding energy, in such a way that the identification of donor pairs in this regime is easier. The charging energy also provides an important independent confirmation of the donor interaction, since it is larger than that of a single donor D$^-$ state. This regime may be adopted for few dopant single electron transistors, leading to an enhanced charging energy and therefore to a more stable operation at room temperature.

\subsubsection{Design of quantum devices.}

Silicon quantum devices take advantage of the quantum behavior of electrons in this semiconductor for specific tasks. Targetting parameters
determined by quantum mechanics is a very challenging task. Purely quantum mechanical quantities, such as
tunnel coupling and Heisenberg exchange coupling, depend explicitly on the wavefunction overlap and are therefore extremely sensitive to the asymptotic decay tails of the orbitals.

\begin{figure}
\leavevmode
\includegraphics[clip,width=0.48\textwidth]{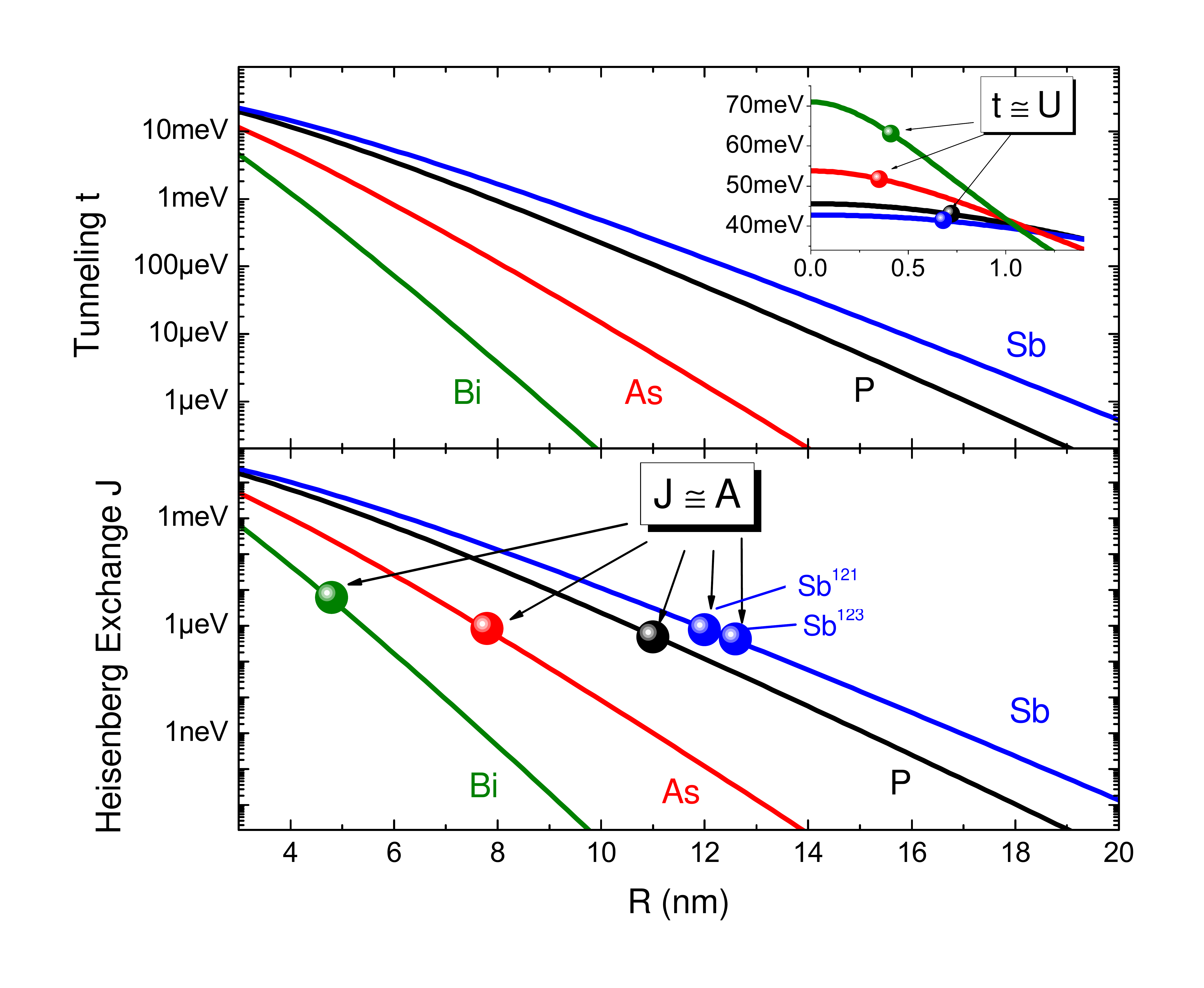}
\caption{Quantum interactions for different impurity species. The tunnel rate is plotted according to the single orbital LCAO simplified expression in Table~\ref{tab:simple}. The data points
for which $t=U$ are marked, adopting the values of $U$ listed in Table~\ref{tab:simple}. The exchange coupling is obtained from the Hubbard model, with the expression also shown in Table~\ref{tab:simple}. The marked values are the data points for which $J=A$, where the hyperfine coupling $A$ is taken from Ref.~[\onlinecite{Feher1959}]. The values of the nuclear spin $I$ and the hyperfine coupling for P, As, Sb$^{121}$, Sb$^{123}$ and Bi are, respectively, $I=$$1/2$, $3/2$, $5/2$, $7/2$ and $9/2$; and $A=$117.5 MHz, 198.3  MHz, 186.8  MHz, 101.5 MHz and 1.4754 GHz.}
\label{fig6}
\end{figure}

Our non-perturbative approach  provides potentials and wavefunctions tailored for each chemical species of the group V substitutional donors. In Fig.~\ref{fig6} we show (a) the tunnel coupling within single orbital LCAO theory and (b) the Heisenberg exchange coupling calculated within the Hubbard model.
Both differ significantly from the results obtained using the KL wavefunction. The KL wavefunction overestimates the real tunnel and exchange
coupling since it does not account for the effect of the valley-orbit coupling in shrinking the envelope function.

The simplified expressions in Sec.~\ref{simple}, on the other hand, involve some approximations that overestimate and some that underestimate the tunnel and exchange couplings.
For instance, taking a spherical effective mass of 0.3$m_e$ leads to a smaller wavefunction overlap compared to the transverse effective mass. On the other hand, disregarding valley
interference and the wavefunction size dependence on the interdonor distance $R$ leads to an increase in the overlap. We expect the net effect of these approximations to have
a lesser negative impact than the approximations in the KL theory, besides providing a path to differentiate and compare impurity chemical species.
We highlight particularly important values of the tunnel and the exchange coupling.

A modest impurity concentration generates impurity bands that are separated by the Mott gap due to the onsite electron-electron repulsion $U$. When the tunnel coupling -- which determines the impurity band width -- becomes comparable to $U$, the system becomes a metal at half-filling. This is the Mott metal-insulator transition. The detailed interdonor distance at which this transition occurs for each impurity is hard to predict -- strong electronic screening is expected at the phase transition. We highlight the distance at which the unscreened repulsion $U$ matches the tunnel coupling $t$.

Another characteristic interdonor distance is that at which the exchange coupling becomes comparable to the nuclear hyperfine coupling. At these distances,
the interaction between neighboring nuclei mediated through its electrons becomes optimally feasible~[\onlinecite{kalra2014}]. This distance is very different amongst different species, as
observed in Fig.~\ref{fig6}. Moreover, this value is markedly different from the hydrogenic estimate adopted originally by Kane~[\onlinecite{kane}], which would set this distance at
$R=23$ nm independently of the chemical species. 

On the other hand, the possibility of shuttling a single electron across the donor pair through a time dependent electric field (in analogy with the quantum operation of electrostatic
quantum dots) is most efficient at $R > 20$nm, for which the tunnel rate is in the range $t/h =$~10 kHz--10~MHz.

\section{Summary and conclusions}
\label{sec:summary}
 
Current progress in single or few dopants characterization~[\onlinecite{rmpreview}] requires higher degree of sophistication where not only the nature (donor or acceptor) but actually identifying the atomic species of each impurity should be accessible to experiment. It is thus desirable to develop simple and reliable theories underlying the physics of such systems. We have presented a comprehensive study of donors in Si, which includes a central cell corrected potential obtained from consistency requirements with experiment. A single species-dependent parameter $r_{cc}$ reproduces data for neutral P, As, Sb and Bi levels within the 1s sixfold lower energy manifold in Si. The parameter $r_{cc}$ characterizes the range of the correction potential, and suffices to differentiate each species and respective potential.

We verify by experimental comparison (when available, otherwise by plausibility arguments) that the proposed central cell corrected potential for a given species in Si  (D$^0)$ is transferrable to other contexts of this species in Si, such as different charge states (D$^-$) and/or other atomic arrangements, {\textit e.g.} donor pairs (D$_2^+$ and D$_2^0$). All approximations involved in different levels of the study are exposed and the range of validity or limitations are explicit. More realistic studies exist in the literature, at the cost of treating only specific centers. For example, while we assume isotropic envelopes,   a recent study~[\onlinecite{klymenko2014}] preserves the Si band anisotropy effects in an analysis restricted to D$_2^+$ (one-electron donor pair ions), with which our results are in fair agreement.

Among the calculated properties, the singlet-triplet difference $\Delta_{\rm ST}$ [see Fig.~5(d)] in the neutral donor pair deserves special attention for donor-based qubits applications. This parameter may be identified with the exchange coupling between the electrons, which is the basic two-qubit entanglement mechanism in the first Si-based quantum processing proposal by Kane~[\onlinecite{kane}]. We obtain the spatial range of the wavefunctions, quantified by $a_{cc}$,  significantly contracted relative to previous estimates based on the screened point-charge potential. One implication is that sizeable  exchange coupling between electrons bound to a donor pair requires interdonor distances significantly smaller than previously expected and attempted. This is probably one of the reasons why exchange-coupled pairs were only reported very recently for As~[\onlinecite{Gonzalez-Zalba2014}] and P~[\onlinecite{dehollain2014}]
in Si. In both cases the models detailed here were instrumental to understand the reported results.

{\it Acknowledgements}. The authors thank M. Friesen and M. Eriksson for fruitful discussions. AS, AB, and BK performed this work as part of the Brazilian National Institute for Science and Technology on Quantum Information and also acknowledge partial support from the Brazilian agencies FAPERJ, CNPq, CAPES. AS also acknowledges the William F. Vilas Trust for financial support. MJC acknowledges support from MINECO-Spain through grant FIS2012-33521. AS, BK and MJC acknowledge support from a bilateral CNPq (Brazil)- CSIC (Spain) grant.


\begin{thebibliography}{99}
\bibitem{foot} For the single donor problem it is convenient to take the donor position as the coordinate origin ${\mathbf r}=0$.
\end{thebibliography}
\end{document}